\documentclass[%
 reprint,
 amsmath,amssymb,
 aps,
]{revtex4-1}

\usepackage{graphicx}% Include figure files
\usepackage{dcolumn}% Align table columns on decimal point
\usepackage{bm}% bold math
\usepackage{dblfloatfix}
\begin{document}

%\preprint{APS/123-QED}

\title{Measurement of $^{34}$S($^{3}$He,p)$^{36}$Cl cross sections for early solar system nuclide enrichment}% Force line breaks with \\
%\thanks{A footnote to the article title}%

\author{Tyler Anderson}
\email{tander15@nd.edu}
\author{Michael Skulski}%
\author{Lauren Callahan}
\author{Adam Clark}
\author{Austin Nelson}
\author{Philippe Collon}

\affiliation{%
 University of Notre Dame\\
}%

\author{Greg Chmiel$^{1}$, Tom Woodruff$^{1}$, Marc Caffee$^{1,2}$}
\affiliation{
 $^{1}$Department of Physics and Astronomy/PRIME Lab, Purdue University
\\
$^{2}$Department of Earth, Atmospheric, and Planetary Sciences, Purdue University}%

\date{\today}% It is always \today, today,
             %  but any date may be explicitly specified

\begin{abstract}
%This is general and still remains. Though I should re-write it somehow?
Isotopic studies of meteorites have provided ample evidence for the presence of short-lived radionuclides (SLRs) with half-lives of less than 100 Myr at the time of the formation of the solar system. The origins of all known SLRs is heavily debated and remains uncertain, but the plausible scenarios can be broadly separated into either local production or outside injection of stellar nucleosynthesis products. The SLR production models are limited in part by reliance on nuclear theory for modeling reactions that lack experimental measurements. Reducing uncertainty on critical reaction cross sections can both enable more precise predictions and provide constraints on physical processes and environments in the early solar system. This goal led to the start of a campaign for measuring production cross sections for the SLR $^{36}$Cl, where Bowers et al. found higher cross sections for the ${}^{33}$S($\alpha$,p)$^{36}$Cl reaction than were predicted by Hauser-Feshbach based nuclear reaction codes TALYS and NON-SMOKER. This prompted re-measurement of the reaction at five new energies within the energy range originally studied, resulting in data slightly above but in agreement with TALYS. Following this, efforts began to measure cross sections for the next most significant reaction for $^{36}$Cl production, $^{34}$S($^{3}$He,p)$^{36}$Cl. Activations were performed to produce 9 samples between 1.11 MeV/nucleon and 2.36 MeV/nucleon. These samples were subsequently measured with accelerator mass spectrometry at two labs. The resulting data suggest a sharper-than-expected rise in cross sections with energy, with peak cross sections up to 30\% higher than predictions from TALYS.

\end{abstract}

\pacs{Valid PACS appear here}% PACS, the Physics and Astronomy
                             % Classification Scheme.
%\keywords{Suggested keywords}%Use showkeys class option if keyword
                              %display desired
\maketitle

%\tableofcontents

\section{\label{sec:intro}Introduction}

Studies performed on meteoritic material have presented evidence for the existence of a large family of short-lived radionuclides (SLRs), nuclei whose half-lives are far shorter than the 4.5 billion year age of the solar system. SLRs are useful as chronometers for astrophysical processes occurring within the solar system, with different isotopic abundances suggesting differing production environments and scenarios \cite{ess_slrs}. Since the first discovery of extinct $^{129}$I from meteoritic excesses of $^{129}$Xe, \cite{Reynolds_1960} many more relic decay products have been found from new isotopic studies on chondrules and Ca-Al-rich inclusions found in carbonaceous chondrite meteorites. Some of the solids containing these decay products did not experience re-melting after their incorporation into the parent meteorite bodies, so they and the isotopic abundances contained within them have been preserved. Others that did experience melting lost their heterogeneity through mixing, but maintained correlations between isotopic abundances, as seen with $^{10}$Be and $^{9}$Be. These studies have presented evidence for an expanding family of SLRs -- such as $^{7}$Be, $^{10}$Be, $^{36}$Cl, $^{41}$Ca, or $^{53}$Mn -- by making correlations between excesses in their decay products relative to their stable isotopes \cite{cai1,cai2,cai3,cai4}.  \\
\indent ${}^{36}$Cl ($t_{1/2} = 0.301$ Myr) is one of three SLRs, along with ${}^{26}$Al, $^{10}$Be, and potentially ${}^{60}$Fe, that has a measured abundance above predictions from galactic steady-state enrichment. This higher abundance suggests the involvement of additional methods of nucleosynthesis in the early solar system\cite{overabundance}.\\
\indent The possible explanations for these overabundances can be categorized as either local production via irradiation processes around the proto-Sun \cite{xwind} or outside injection of stellar nucleosynthetic products \cite{ess_nuc1,ess_nuc2,ess_nuc3}. Early solar system solids with evidence of extinct $^7$Be and $^{10}$Be -- both of which are known only to be produced in spallation reactions -- show that intense irradiation processes took place around the time of their formation \cite{7Be_10Be,Mckeegan}. The presence of $^{60}$Fe -- a nuclide produced only in stellar environments -- also shows that some amount of material must have been injected from outside the solar system.  Neither of these scenarios has yet been able to produce models capable of explaining all SLR abundances. This is in part due to the large number of related reactions for which no experimental measurements exist, causing reliance on predictions from nuclear theory\cite{ess_slrs,slr_abund}. Hauser-Feshbach based calculations are commonly employed to predict relevant cross sections, for example in the nuclear reaction code TALYS \cite{talys1,talys2}, but are generally accepted to have uncertainties up to a factor of three \cite{ess_err}. To reduce this uncertainty and increase the predictive power of early solar system models, a measurement campaign for $^{36}$Cl producing reactions was started.\\
\indent Bowers et al. performed the first measurements toward this goal by measuring cross sections for the ${}^{33}$S($\alpha$,p)${}^{36}$Cl reaction between 0.70 and 2.42 MeV/nucleon \cite{bowers}. The resulting data showed a systematic under-prediction of cross sections by the Hauser-Feshbach codes TALYS and NON-SMOKER, the irregularity of which was highlighted in a following paper by Mohr \cite{mohr}. In response, the reaction was re-measured at 5 points in the energy range where experimental data and theory were most discrepant between 0.78 MeV/nucleon and 1.52 MeV/nucleon, with a special focus paid to the procedures followed by Bowers et al. to rule out any experimental errors \cite{My_36Cl}. The new samples were produced shortly before a scheduled maintenance shutdown, so they were measured at Purdue's Rare Isotope Measurement Laboratory (PRIME Lab). The re-measured data agreed with the predictions from the nuclear reaction code TALYS \cite{talys1,talys2}.\\
\indent This work represents the next step in the $^{36}$Cl campaign with measurements of ${}^{34}$S($^3$He,p)${}^{36}$Cl across as wide an energy range as was experimentally feasible given experimental equipment limitations and predicted reaction cross sections. The details of the experiment are nearly identical to those for the re-measurement of the $^{33}$S($\alpha$,p)$^{36}$Cl reaction discussed in \cite{My_36Cl} and are broken into three parts: activations, extraction chemistry, and sample measurement with Accelerator Mass Spectrometry (AMS). \\

\section{\label{sec:exp}Experimental Procedure}

\subsection{\label{sec:activations}Activations}

\begin{figure}[tbp]
    \includegraphics[width=0.48\textwidth]{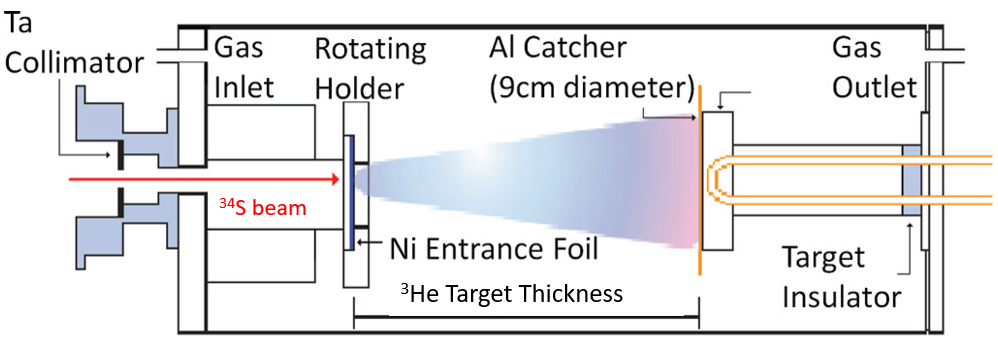}
    \caption{Reproduced from \protect\cite{bowers}, a schematic of the gas cell used for the activations. A 9cm diameter Al foil was used to catch the forward-recoiled ${}^{36}$Cl atoms.}
    \label{figure:gas_cell}
\end{figure}

\begin{figure}[tpb]
    \includegraphics[width=0.48\textwidth]{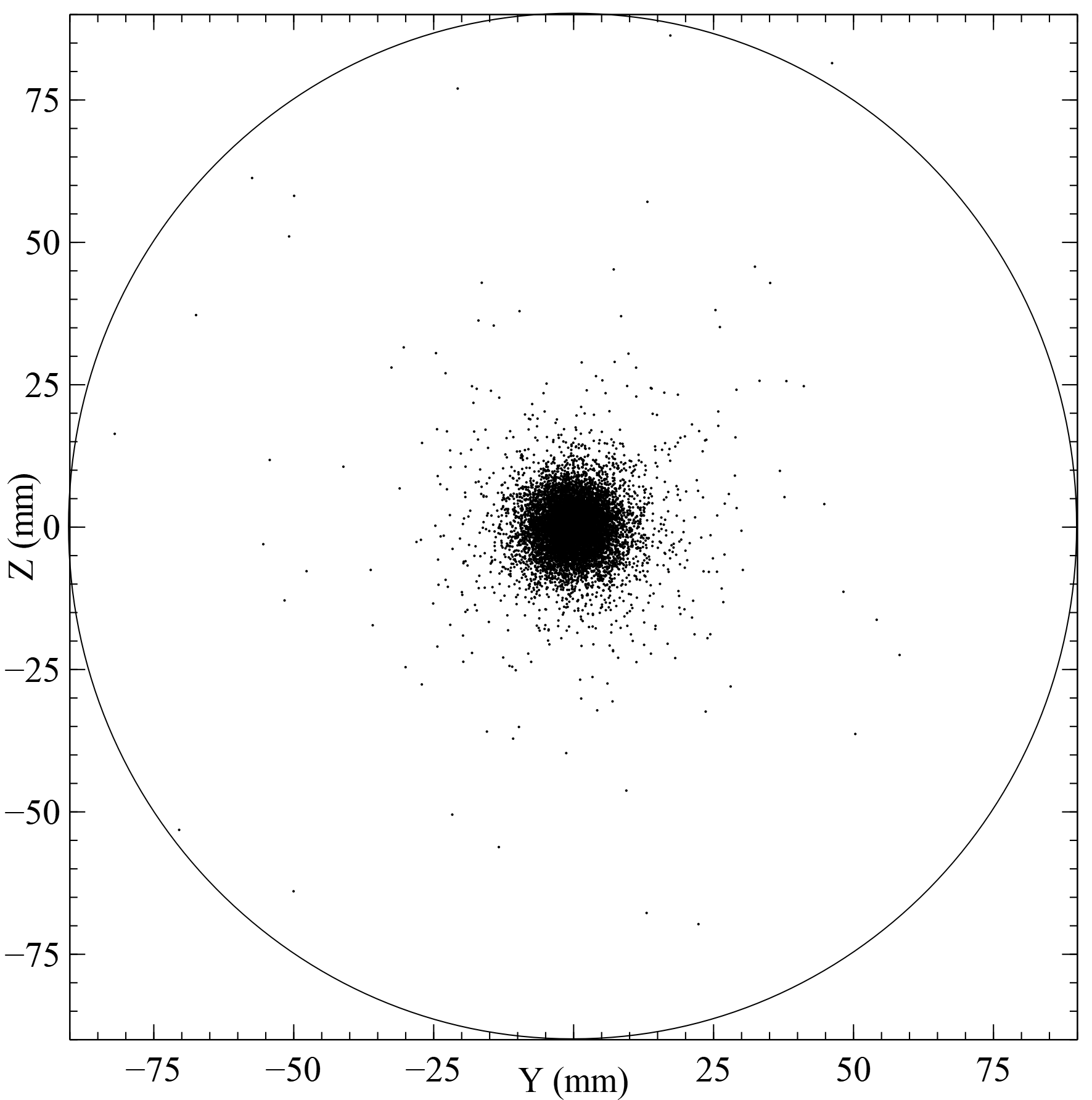}
    \caption{The results from a SRIM simulation of $10^4$ ions of ${}^{34}$S passing through the ${}^{3}$He filled gas cell for the lowest energy sample (68 MeV). 99.9\% of recoils are caught within the 9 cm diameter  }
    \label{figure:TRIM}
\end{figure}

The activations to produce $^{36}$Cl took place at the Nuclear Science Laboratory at the University of Notre Dame (NSL). Nine ${}^{36}$Cl samples were created via ${}^{34}$S($^3$He,p) in inverse kinematics at energies between 1.11 MeV/nucleon and 2.36 MeV/nucleon. An FeS cathode was used to produce ${}^{34}$S beam from an MC-SNICS, which was accelerated by an FN Tandem Van de Graaff into a $^3$He filled gas cell, shown in Figure \ref{figure:gas_cell}. Any created $^{36}$Cl atoms were forward-recoiled and implanted into an Al catcher foil at the back of the cell. Aside from minor changes made to re-capture the target $^3$He gas and minimize its losses, this method has been successfully used to produce samples from the $\alpha$-induced reactions ${}^{40}$Ca($\alpha$,$\gamma$)${}^{44}$Ti \cite{danthesis} and $^{33}$S($\alpha$,p)$^{36}$Cl \cite{bowers,My_36Cl}. The reaction energies were chosen based upon the predicted cross sections using the default parameters of TALYS, the energy loss through the gas cell as calculated by SRIM\cite{srim}, and the voltage limitations of the accelerator.\\
\indent The gas cell was attached as an end-cap to the beampipe and used a 2.5 $\mu$m thick Ni foil as a window to separate its volume from vacuum. The front of the cell is a large insulator containing two collimators, electrically isolating the cell and allowing collimator current to be read to aid tuning. Beam was tuned into the cell with the entrance window removed and an insulator installed between the catcher foil holder and the back of the cell to allow reading the current on target. During activations, the entrance window was rotated off axis via an electrically isolated external motor to reduce degradation over time from the $^{34}$S beam. Additionally, a 0.25 mm thick Al catcher foil was mounted to a brass holder and the insulator used for tuning was removed so all current incident on the gas cell could be collected and integrated, allowing for determination of the number of incident $^{34}$S ions. The catcher foil holder was continuously cooled during activations via compressed air forced through a u-shaped bend of tubing exiting the back of the gas cell. $^{3}$He was flowed into the gas cell using an automated gas handling system which monitored and maintained the target gas pressure. To reduce $^3$He use, the target gas was not re-circulated, but additional $^3$He was flowed to replace losses to vacuum through the entrance foil, negligibly changing the average pressure over the length of an activation. Once the $^{34}$S beam passed through the entrance window, it could react at any point along the 24 cm path between entrance and catcher foils before implanting in the catcher foil, giving each reaction an integrated energy range.\\
\indent SRIM was used to calculate energy lost through the Ni entrance foil and ${}^{3}$He gas, shown in Table \ref{table:de/dx}. The energy range for each activation is $E_{low}$ to $E_{high}$, defined by
$$E_{high} = E_{foil} + FWHM/2,$$
$$E_{low} = E_{gas} - FWHM/2,$$
\noindent where $E_{foil}$ and $E_{gas}$ are the centroid of the beam energy after the entrance foil and after the $^{3}$He gas, respectively, and $FWHM$ is the full width at half maximum of the beam energy distribution.\\
\begin{table}[htbp]
\centering
\begin{tabular}{|c|c|c|c|c|c|c|c|} 
 \hline
 Sample & $E_{i}$ & $E_{foil}$ & $E_{gas}$ & $FWHM$ & $E_{high}$ & $E_{low}$ & $\Delta E$ \\ 
 \hline
 $^{34}$S-1 & 68    & 40.5 & 38.2 & 0.76 & 40.8 & 37.8 & 3.07 \\
 $^{34}$S-2 & 71.1  & 43.8 & 41.6 & 0.78 & 44.2 & 41.1 & 3.06 \\
 $^{34}$S-3 & 76.0  & 49.1 & 46.9 & 0.83 & 49.5 & 46.5 & 3.02 \\
 $^{34}$S-4 & 79.0  & 52.3 & 50.2 & 0.77 & 52.8 & 49.8 & 2.94 \\
 $^{34}$S-5 & 83.25 & 56.9 & 54.8 & 0.71 & 57.2 & 54.4 & 2.83 \\  
 $^{34}$S-6 & 86.9  & 60.8 & 58.8 & 0.77 & 61.2 & 58.4 & 2.85 \\
 $^{34}$S-7 & 90.0  & 64.2 & 62.1 & 0.70 & 64.5 & 61.8 & 2.72 \\
 $^{34}$S-8 & 95.0  & 69.5 & 67.4 & 0.77 & 69.9 & 67.0 & 2.93 \\ 
 $^{34}$S-9 & 104.5 & 79.8 & 77.9 & 0.81 & 80.1 & 77.4 & 2.68 \\ 
 \hline
\end{tabular}
\caption{Information on energy loss of the ${}^{34}$S beam as it passed through the gas cell. Shown for each sample are, (1) $E_i$ is the incident beam energy before passing through the Ni foil, (2) $E_{foil}$ is the mean energy after the Ni foil, (3) $E_{gas}$ is the mean energy after passing through the ${}^{3}$He gas, and (4) $FWHM$ is the full width at half maximum of the beam energy distribution after the He gas. (5) $E_{high}$ and (6) $E_{low}$ are the high and low bounds in reaction energy, calculated as described in section \protect\ref{sec:activations}. (7) $\Delta E$ is the energy range for each reaction energy. All values listed are measured in MeV.}
\label{table:de/dx}
\end{table}
\begin{table}[!bp]
\centering
\small
\setlength\tabcolsep{2pt}
\begin{tabular}{|c|c|c|c|c|c|c|} 
 \hline
Sample	&	Activation 	&	I$_{avg}$	&	$^{34}$S Charge	&	N$_{^{34}S}$  	&	N$_{target}$  	  \\ 
	&	 Length (hr)	&	(nA)	&	State	&	(10$^{15}$) 	&	 (10$^{22}$/m$^2$)	 \\
 \hline
$^{34}$S-1 &	11.7	&	870.1	&	7$^+$	&	32.68	&	7.9	\\
$^{34}$S-2 &	3.65	&	564.5	&	8$^+$	&	5.79	&	7.9	\\
$^{34}$S-3 &	1.48	&	537.6	&	7$^+$	&	2.56	&	7.9	\\
$^{34}$S-4 &	1.88	&	295.0  	&	9$^+$	&	1.39	&	7.9	\\
$^{34}$S-5 &	0.33	&	841.7	&	8$^+$	&	0.79	&	7.9	\\
$^{34}$S-6 &	5.23	&	40.0 	&	10$^+$	&	0.47	&	7.9	\\
$^{34}$S-7 &	0.27	&	435.9	&	9$^+$	&	0.29	&	7.9	\\
$^{34}$S-8 &	0.27	&	291.3	&	9$^+$	&	0.19	&	8.7	\\
$^{34}$S-9 &	0.58	&	129.4	&	10$^+$	&	0.17	&	7.9	\\
 \hline
\end{tabular}
\caption{Presented for each sample are (1) the activation length, (2) the average $^{34}$S electrical current incident on the cell, (3) the charge state of the beam, (4) the corresponding number of incident $^{34}$S ions in 10$^{15}$ atoms, and (5) the $^3$He target density in 10$^{22}$ atoms/m$^2$. Sample 8 has a higher target density corresponding to a pressure of 11 Torr due to the high gas pressure in the source bottle for $^3$He resulting in the GHS overshooting the set pressure. Due to the minor increase, instead of taking time and extra losses by pumping out the excess $^3$He, the decision was made to proceed as the minor change in pressure would not negatively impact the experiment.}
\label{table:act}
\end{table}
\indent Recoil simulations performed in SRIM show that more than 99.9\% of recoils are caught within the 9 cm diameter circular cross section of the catcher foil (Figure \ref{figure:TRIM}), and are implanted at least 11 $\mu$m deep, making losses due to sputtering of $^{36}$Cl unlikely. Information about the length of each activation, the incident beam intensity, and target density are shown in Table \ref{table:act}.\\

\subsection{\label{sec:chem}Chemistry}
\indent First, the Al catcher foil was cut into pieces small enough to fit in a 600 ml High-density polyethylene (HDPE) bottle. Each bottle then had a precisely measured amount of stable Cl carrier in solution added which determines the $^{36}$Cl/Cl concentration, shown in Table \ref{table:chem}. It is important that the dissolution reaction is allowed to proceed slowly, because the high temperatures associated with an over-vigorous reaction can lead to preferential boiling off of stable Cl due to $^{36}$Cl still being contained in the as-yet undissolved foil parts, which would lead to an indeterminable sample concentration. To prevent this, 40 g of 18M$\Omega$ de-ionized (DI) H$_2$O is added as a buffer and heat sink. A total of 45 g of HF (49\%) was added, starting with an initial 12 g of HF and adding more in 3 g increments, allowing the reaction to subside after each addition.\\ 
\indent In previous dissolutions, a thick gel determined to be AlF$_2$ formed in some samples, resulting from the foils reacting with HF. This prevented extraction of sample AgCl as the powder was trapped inside the gel. To prevent formation of this gel, an additional 50 ml of DI H$_2$O is added to each bottle for dissolution to dilute the AlF$_2$ and prevent it from precipitating. \\
\indent The Cl is precipitated as AgCl with the addition of AgNO$_3$ in excess, such that every Cl atom has an Ag atom to pair with. The samples were then spun in a centrifuge to compact the AgCl into a pellet, and the remaining liquid was removed. The AgCl was rinsed by breaking up the pellet, adding DI water, and centrifuging again to re-compact it. The remaining liquid was removed once more before the samples were placed in an oven to dry at around 80 $^{\circ}$C overnight. Sample collection efficiency as determined by the mass yield of AgCl was 87-92\%.\\
\begin{table}[htbp]
\centering
\begin{tabular}{|c|c|c|} 
 \hline
 Sample & $M_{carrier}$ (g) & $N_{Cl}$ ($10^{19}$ atoms) \\ [0.5ex] 
 \hline
$^{34}$S-1	&	4.0643	&	6.99	\\
$^{34}$S-2	&	4.1274	&	7.10	\\
$^{34}$S-3	&	4.1855	&	7.20	\\
$^{34}$S-4	&	4.0288	&	6.93	\\
$^{34}$S-5	&	4.0053	&	6.89	\\
$^{34}$S-6	&	4.1630	&	7.16	\\
$^{34}$S-7	&	4.0165	&	6.91	\\
$^{34}$S-8	&	4.0505	&	6.97	\\
$^{34}$S-9	&	4.0390	&	6.95	\\
Chem Blank 	&	2.0908	&	3.70	\\
 
 \hline
\end{tabular}
\caption{The amount of stable Cl carrier added (with a concentration of 1.013mg/g Cl), and the equivalent number of Cl atoms added to each sample.}
\label{table:chem}
\end{table}

\subsection{\label{sec:ams}Accelerator Mass Spectrometry}

\begin{figure}[htbp]
    \includegraphics[width=0.48\textwidth]{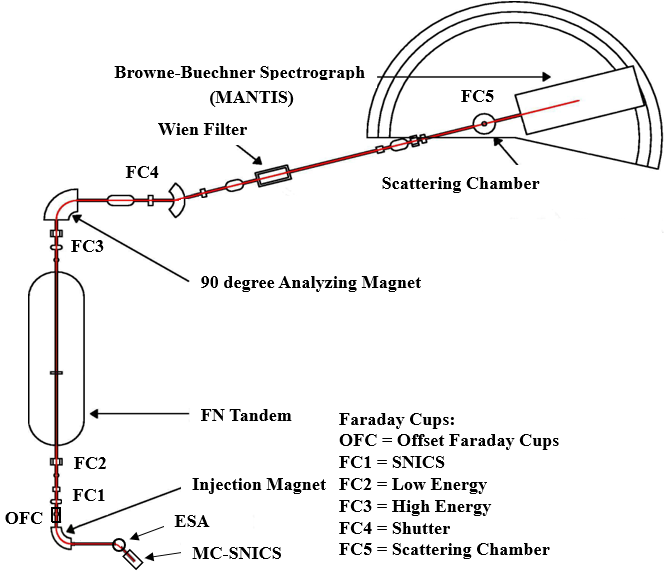}
    \caption{The FN Tandem accelerator at the NSL and the AMS beamline are shown along with key elements. The AMS beamline underwent a redesign during a scheduled accelerator shutdown to improve transmission, and a high energy offset Faraday cup was installed after the analyzing magnet to improve AMS measurements.}
    \label{figure:NSL}
\end{figure}

\indent The first AMS measurements were performed at the NSL (see Figure \ref{figure:NSL} for a schematic layout) in the weeks prior to a scheduled shutdown. The goals of the shutdown included installation of new experimental equipment and maintenance to repair instabilities in the FN Tandem accelerator and critical beam focusing elements. As a result, the AMS data from this period was subject to significantly higher uncertainties than typical. Given these issues and prior contested AMS results for $^{34}$S($\alpha$,p), the remaining sample material was measured at PRIME Lab to verify the NSL AMS data. During the course of the shutdown, a high energy offset Faraday cup was installed to aid AMS measurements and the AMS beamline underwent a re-design to improve transmission.\\
\indent The same method for sample preparation was used at both the NSL and PRIME Lab, and is identical to the method used in \cite{My_36Cl}. To prepare a sample for use in a sputtering ion source, the extracted AgCl powder was gently pressed into the face of AgBr-packed cathodes which were warmed on a hot plate to drive away moisture. Due to its low sulfur content, use of AgBr as a back-packing material reduces production of the main isobaric contaminant for $^{36}$Cl measurement, $^{36}$S. Both labs accelerated the low energy Cl ion beam extracted from each sample with an FN Tandem to produce $^{36}$Cl beams with energies of 74.3 MeV at the NSL and 84.3 MeV at PRIME Lab.\\
\indent At the NSL, ${}^{35}$Cl and ${}^{37}$Cl beams were continuously measured in offset Faraday cups before acceleration to allow final determination of sample concentrations. Isotopic selection was performed by a high energy analyzing magnet and a Wien filter, while isobaric separation of ${}^{36}$Cl from ${}^{36}$S was performed with a 90 degree Browne-Buchner spectrograph filled with 2.7 Torr of N$_2$ gas. Located at the exit of the spectrograph magnet are a parallel grid avalanche counter and an ionization chamber -- filled with 3 Torr and 9 Torr of isobutane, respectively -- which are used together for particle identification. Two standard materials obtained from PRIME Lab with ${}^{36}$Cl/Cl concentrations of 4.16 $\times$ 10$^{-11}$ and 4.42 $\times$ 10$^{-12}$ were used to normalize the AMS measurements, and the chemistry blank was used to determine appropriate background subtraction.\\
\indent At PRIME Lab, after acceleration and beam analysis are a pair of high energy offset Faraday cups which were used to measure chopped $^{35}$Cl and $^{37}$Cl currents while the mass 36 beam was passed through 3 consecutive Wien filters for isotopic separation. Isobaric separation was performed using a 135 degree magnet filled with 4 Torr of N$_2$ gas. At the exit of the magnet is an ionization chamber filled with 85 Torr of P-10 used for particle identification.\\

%\indent The AMS measurements were normalized to the $^{36}$Cl standard material KNSTD500. The procedures for the development of these standards are given in Sharma et al \cite{sharma}. The standards used for this specific measurement were also those used for a recent calibration of $^{36}$Cl production rates. These results are described in Marrero et al \cite{marrero}. Five secondary standards were also measured. The results for these secondary standards, when compared with nominal values, yielded a chi-squared value of 7.03. With 4 degrees of freedom this gives a p-value of 0.13. These measurements were not significantly different from the nominal values at the $\alpha$=.10 level. Machine blanks measured yielded a value of 5.2 $\pm$ 0.4 $\times$ $10^{-15}$ $^{36}$Cl/Cl, not significantly different from the historical value of 5 $\times$ $10^{-15}$ $^{36}$Cl/Cl. Taken together the measurements of these standards indicates a linear response of the AMS system. We can additionally conclude that there were no sources of contamination from the ion source that could have impacted these measurements. \\
\indent The measured $^{36}$Cl/Cl concentrations and associated errors are shown in Table \ref{table:34S_concentrations} for both the NSL and PRIME Lab.\\

\begin{table}[htbp]
\centering
\begin{tabular}{|c|c|c|c|c|c|} 
\hline
    Sample & $E_{low} - E_{high}$ & NSL & error & PRIME & error \\ 
 & (MeV/A)& $(10^{-15})$ & $(10^{-15})$ & $(10^{-15})$ & $(10^{-15})$ \\
 \hline

$^{34}$S-1	&	1.11-1.20	&	2196	&	370	&	1912	&	36 \\
$^{34}$S-2 	&	1.21-1.30	&	1392	&	299	&	1479	&	19 \\
$^{34}$S-3 	&	1.37-1.46	&	2073	&	365	&	1865	&	20 \\
$^{34}$S-4 	&	1.47-1.55	&	1967	&	338	&	1885	&	22 \\
$^{34}$S-5 	&	1.60-1.68	&	2299	&	294	&	2067	&	22 \\
$^{34}$S-6	&	1.72-1.80	&	2537	&	371	&	2231	&	25 \\
$^{34}$S-7 	&	1.82-1.90	&	1307	&	82	&	1013	&	14 \\
$^{34}$S-8 	&	1.97-2.06	&	1727	&	372	&	1565	&	24 \\
$^{34}$S-9 	&	2.28-2.36	&	1748	&	294	&	1691	&	33 \\
 \hline
\end{tabular}
\caption{Shown for each sample are, (1) the reaction energy ranges as determined as described in Section \protect\ref{sec:activations}, (2) the measured ${}^{36}$Cl/Cl concentration at the NSL and (3) the associated error, (4) the measured ${}^{36}$Cl/Cl concentration at the PRIME Lab and (5) the associated error. The NSL AMS was impacted by significant equipment instabilities and the error for the measurements is significantly higher than typical values.}
\label{table:34S_concentrations}
\end{table}

\section{\label{sec:results}Results}
\begin{figure}[htbp]
    \includegraphics[width=0.48\textwidth]{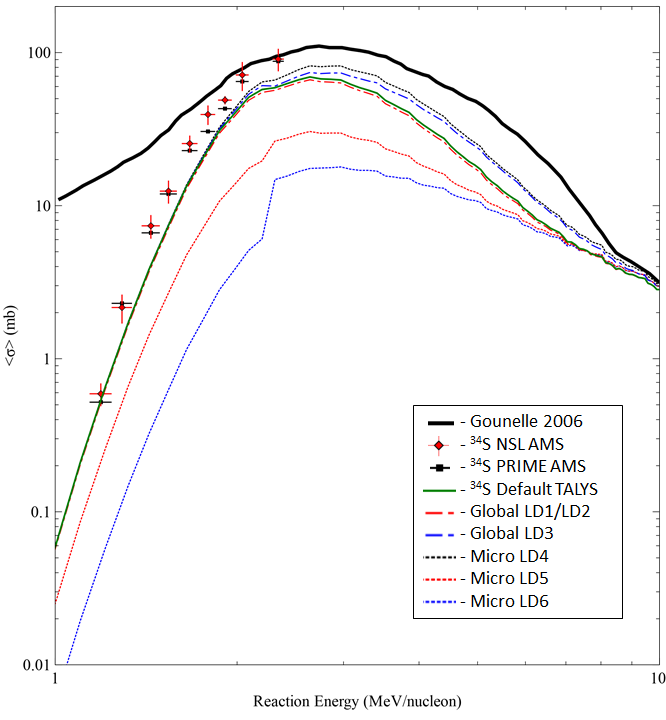}
    \caption{Experimental data from AMS measurements are compared to predictions from the code TALYS and the adopted cross sections by Gounelle et al \cite{gounelle}. The red points are the calculated integrated cross sections from NSL AMS data, and the black points are from PRIME Lab AMS data. In solid black are the cross sections used by Gounelle, in solid green is the predicted cross section curve with TAYLS's default parameters, and the rest of the curves are results using one of the six different level density (LD) models available in the code. The dash-dotted lines are results using global/phenomenological models, with LD1 and LD2 (which produce the same results across this energy range) shown in red, and LD3 in blue. The dotted lines are results using microscopic level density models with LD4 in black, LD5 in red, and LD6 in blue.}
    \label{figure:results}
\end{figure}

The cross sections were calculated in the same way as \cite{My_36Cl,bowers} using
\begin{equation}
    <\sigma> = \frac{N_{36Cl}}{N_{34} \times N_{T}},
\end{equation}
where $N_{36Cl}$ is the number of ${}^{36}$Cl atoms calculated in the sample, $N_{34}$ is the total number of incident ${}^{34}$S ions for an activation, and $N_{T}$ is the areal density of ${}^{3}$He target atoms. $N_T$ is calculated with
\begin{equation}
 N_T=\rho_{atm}\frac{P}{P_{atm}}\frac{N_A}{M_{He}}d   
\end{equation}

\noindent where $N_T$ is in units of target nuclei/cm$^2$, $\rho_{atm}$ (=125.3 g/$m^3$) is the density of ${}^3$He at atmospheric pressure, and $P$ and $P_{atm}$ are gas cell and atmospheric pressures, respectively. $N_A$ is Avogadro's constant, $M_{He}$ is the atomic mass of helium-3 (=3.0160 g/mol), and $d$ (=24 cm) is the distance between the Ni entrance foil and Al catcher foil in the gas cell.\\ 
\indent The number of $^{36}$Cl atoms present in each sample as determined by AMS, along with their respective reaction energy ranges and calculated integrated cross sections, are listed in Table \ref{table:results}. The uncertainties in the measurements involved in calculating the reported cross sections are listed in Table \ref{table:error}. These cross sections are plotted in Figure \ref{figure:results} with the predicted cross sections from default TALYS, as well as the predictions resulting from using each of the six different level density models available in TALYS, discussed in Section \ref{sec:disc} . 

\begin{table}[htbp]
\centering
\begin{tabular}{|c|c|c|c|c|c|} 
 \hline
 Sample & E$_{low}$ - E$_{high}$  & \multicolumn{2}{c|}{N$_{{}^{36}Cl}$ (10$^8$ atoms)} &\multicolumn{2}{c|}{$<\sigma>$ (mb)}   \\ 
 & (MeV/A) & (NSL) & (PRIME) & (NSL) & (PRIME)\\
 \hline
 $^{34}$S-1 & 1.11 - 1.20 & 1.54 & 1.34 & 0.6(1) & 0.52(1) \\
 $^{34}$S-2 & 1.21 - 1.30 & 0.99 & 1.05 & 2.2(4) & 2.30(3) \\ 
 $^{34}$S-3 & 1.37 - 1.46 & 1.49 & 1.34 & 7.4(1.3) & 6.64(7) \\
 $^{34}$S-4 & 1.47 - 1.55 & 1.36 & 1.31 & 12(2) & 11.9(1) \\
 $^{34}$S-5 & 1.60 - 1.68 & 1.58 & 1.42 & 25(3) & 22.9(2) \\
 $^{34}$S-6 & 1.72 - 1.80 & 1.82 & 1.60 & 39(5.7) & 30.5(3) \\
 $^{34}$S-7 & 1.82 - 1.90 & 0.90 & 0.7 & 49(3) & 43.0(6) \\
 $^{34}$S-8 & 1.97 - 2.06 & 1.20 & 1.09 & 71(15) & 64.7(1.0) \\ 
 $^{34}$S-9 & 2.28 - 2.36 & 1.21 & 1.18 & 91(15) & 87.7(1.7) \\
\hline
\end{tabular}
\caption{Shown for each sample are, (1) the reaction energy ranges as determined as described in Section \protect\ref{sec:activations}, (2) the number of ${}^{36}$Cl atoms produced as determined by the concentrations measured using AMS from both labs, and (3) the calculated integrated cross section from each lab's AMS data.}
\label{table:results}
\end{table}

\begin{table}[htbp]
\centering
Measurement error budget
\begin{tabular}{|c|c|} 
\hline

 Incident ${}^{34}$S ions ($N_{34}$) & 2\% \\ 
 Stable Cl carrier atoms ($N_{Cl}$) & 1\%   \\
 ${}^3$He target density & 2\%   \\
 ${}^{36}$Cl/Cl (NSL) & 6-22\%  \\
 ${}^{36}$Cl/Cl (PRIME) & 1-2\%  \\

 \hline
\end{tabular}
\caption{A summary of uncertainties used for the different measurements.}
\label{table:error}
\end{table}
\section{\label{sec:disc}Discussion}
\indent The integrated cross sections presented here for $^{34}$S($^3$He,p) represent values 20-30\% higher than the calculations produced using the default parameters of TALYS for all but the lowest energy point. This suggests that, while the general shape of the cross section curve is correct, the predictions peak at too low of an energy. The results observed for $^{33}$S($\alpha$,p) \cite{My_36Cl} showed an apparent disagreement with the trend of Hauser-Feshbach codes slightly over-predicting cross sections for $\alpha$-induced reactions \cite{mohr}, and while this reaction was $^3$He-induced it may still be similarly categorized. To ensure confidence in the experimental results, all sources of systematic error in the measurements from this work will be addressed.\\
\indent The calculated cross sections are impacted by the reliability of the measurements for 1) the target gas pressure, 2) the target gas purity, 3) the number of incident sulfur ions, 4) the amount of stable chlorine carrier present in each sample, and 5) the measured sample concentration. The reasons why each of these did not artificially inflate the cross sections will be addressed in order. 1) The target gas pressure was constantly monitored and was independently verified with a second, external pressure gauge. 2) The $^3$He target gas was previously unused and nominally 99.9\% pure. 3) The electrical isolation of the gas cell was regularly verified, the current integrator used was verified to be accurate with an external current source, and noise on the current integrator was 2\% or below and accounted for. 4) The scale used to determine the masses of chlorine carrier added to each sample was calibrated immediately prior to the measurements, and was confirmed to still be within calibration afterward. Additionally, the foil dissolution chemistry procedure has been verified to preserve the $^{36}$Cl/Cl ratio \cite{My_36Cl}. 5) While the AMS measurements at the NSL were impacted by significant instabilities in critical experimental equipment including the MC-SNICS ion source, FN Tandem accelerator, and high energy analyzing magnet power supply, the measured sample concentrations from the NSL and PRIME Lab agree within error for all samples but 6 and 7. Many NSL measurements were ultimately rejected after the measured ratio of 10$^{-11}$ and 10$^{-12}$ standards was found to be inconsistent, or standard measurements before and after a sample were very different, making accurate normalization of the sample concentrations with the standard material impossible.\\
\indent Similarly, inaccuracies in the energy of the beam at any point in the activation cell could artificially shift the associated energy for each reported cross section. SRIM was used to calculate energy loss and is reported to have a 2\% or less deviation between simulation and experiment at these energies \cite{srim}. Otherwise, reaction energies could be impacted by inaccuracies in 1) the beam energy before entering the gas cell, 2) the thickness of the Ni entrance foil, and 3) the target gas pressure, which was already addressed above. 1) The beam energy selected by the NSL analyzing magnet has been confirmed to be accurate to within $<$ 10 keV through measurements of the $^{27}$Al(p,$\gamma$)$^{28}$Si reaction \cite{hector}. 2) The thicknesses of both un-used and heavily used Ni entrance foils were verified with $\alpha$-spectroscopy to be identical.\\
\indent After all experimental effects that could shift the data were ruled out, a sensitivity study was performed within TALYS to ascertain whether any arrangement of models was capable of more accurately reproducing the data. The models that were varied include the $\alpha$ optical model potential, the proton optical model potential, the level density, and the $\gamma$-strength function. Two sets of model arrangements were used consisting of the available global/phenomenological models and the available microscopic models. The predicted cross sections were entirely dependent on the choice of level density model. The differences resulting from changing optical models and $\gamma$-strength function models were negligible or non-existent. The predictions from the global LD models only begin to diverge significantly above 2 MeV/nucleon, while the microscopic models make very different predictions at low energy before converging around 8 MeV/nucleon. With the exception of the lowest energy point, none of the predictions made with the different LD models agreed with the experimental data.\\

\subsection{Astrophysical Implications}

\indent Observations of young stellar objects (YSOs) have shown stars in their T Tauri stage are subject to large x-ray flare events capable of accelerating nearby nuclei, from protons to $^4$He, to energies up to and above 10 MeV/nucleon, referred to as solar energetic particles (SEPs). Under the x-wind model, SEPs irradiate CAIs and chondrules very close to the sun to produce many of the SLRs that have been observed in meteoritic material \cite{xwind}. The particle fluences from these accelerating flare events are modeled with a power law distribution $\propto E^{-p}$, where $E$ is the particle energy and p is a parameter that varies between 2.7 and 5 to adjust the shape of the energy spectrum. The x-ray flares are divided into gradual events, in which the SEP energy spectrum is shallow with a low p and has a low $^3$He fluence, and impulsive events, in which the SEP energy spectrum is sharp with a higher p, and has a high $^3$He fluence \cite{gounelle,ess_err}.\\
\indent Bowers et al. highlighted the $^{34}$S($^{3}$He,p) reaction as one of the most significant contributors to producing $^{36}$Cl, even under the assumption of lower cross sections across the entire energy range. Additionally, the cross sections adopted by Gounelle et al. \cite{gounelle} for the $^{34}$S($^3$He,p)$^{36}$Cl reaction significantly over-predict cross sections relative to all TALYS models below 10 MeV/nucleon, but appear to capture the experimentally measured peak reaction cross section accurately. Given the total lack of experimental data at the time, the peak cross sections were assumed to have a 50\% uncertainty, whereas all other cross sections away from the peak were assumed to have a factor of 2 uncertainty. The latter assumption appears to break down below 1.7 MeV/nucleon where deviations quickly diverge past a factor of 2, though given the rest of the un-measured energy range, it is not possible to comment on the quality of the assumption above $\approx$ 3 MeV/nucleon. As a result, particularly because of the steep SEP spectrum assumed, meaning a higher fluence of lower energy SEPs were present, the under production of $^{36}$Cl may be further exacerbated. Alternatively, in a model constrained by $^{36}$Cl production alone, producing the inferred initial $^{36}$Cl abundances will lead to all other SLRs being necessarily over-produced under the same circumstances. With the data from this work, this reaction becomes even more critical for an accurate accounting of $^{36}$Cl production.\\
\indent These higher cross sections seem to exacerbate a problem within the x-wind model where co-production of $^{36}$Cl with other SLRs, such as $^{26}$Al and $^{53}$Mn, leads to their overproduction compared to values measured from meteorites. While the x-wind model inherently assumes refractory target material due to the high temperatures present close to the young sun, results from Jacobsen et al. \cite{cai4} suggest an alternative scenario where $^{36}$Cl was produced independently of other SLRs. The proposed environment is at a greater distance from the Sun in a volatile-rich reservoir in the protoplanetary disk, and would occur $>$ 2 Myr after formation of the first solar system solids. This scenario likely benefits further from accurate measurements of $^{34}$S($^{3}$He,p) due to the greater presence of volatile sulfur. To be able to fully determine the origins of $^{36}$Cl in the early solar system, higher energy cross section measurements are important, and as the energy range available at the NSL makes measurements at and past the peak cross section unfeasible, measurements from other labs are needed. Given the possible high inferred irradiation energies ($>$ 10 MeV/nucleon), the disagreement in TALYS predictions between 2 MeV/nucleon and 10 MeV/nucleon, and the under-prediction of cross sections across almost the entire measured energy range, additional measurements in this energy range and above could be of importance not only for early solar system models, but also nuclear theory.\\

\section*{Acknowledgements}
Thanks are extended to PRIME Lab for all of their guidance in developing the chemistry, for measuring our samples at their lab, and their assistance with this article. Additional thanks to Anna Simon for her input in the TALYS sensitivty study, and to Yoav Kashiv for his idea originating this measurement campaign. PRIME Lab personnel acknowledge support from NSF EAR-0844151. This work is supported by National Science Foundation Grant No. NSF PHY-1419765. 

\bibliographystyle{unsrt}
\bibliography{refs}

\end{document}